\documentclass[twocolumn]{aastex63}

\hypersetup{linkcolor=red,citecolor=green,filecolor=cyan,urlcolor=magenta}

\newcommand{\rh}{$r_{h}$}
\newcommand{\cubi}{$C_{U,B,I}$}
\defcitealias{lar11}{L11}
\defcitealias{sav18}{S18}
\defcitealias{mon13}{Mon13}
\defcitealias{ste19}{S19}

\received{}
\revised{}
\accepted{}

\submitjournal{AJ}

\shorttitle{Mixed Populations in M13}
\shortauthors{Smolinski et al.}

\begin{document}
\title{Re-examining the Radial Distributions of M13 Multiple Populations}

\correspondingauthor{Jason Smolinski}
\email{js85@calvin.edu}

\author[0000-0002-2045-7353]{Jason P. Smolinski} 
\affiliation{Department of Physics \& Astronomy, Calvin University, Grand Rapids, MI 49546, USA}

\author{Willem B. Hoogendam}
\affiliation{Department of Physics \& Astronomy, Calvin University, Grand Rapids, MI 49546, USA}

\author[0000-0002-3905-4853]{Alex J. Van Kooten}
\affiliation{Department of Physics \& Astronomy, Calvin University, Grand Rapids, MI 49546, USA}

\author{Peyton Benac}
\affiliation{Harvard-Smithsonian Center for Astrophysics, Cambridge, MA 02139, USA}

\author{Zachary J. Bruce}

\begin{abstract}
We seek to resolve the tension in the literature regarding 
the presence of radially segregated multiple populations in the Galactic 
globular cluster M13. Previous studies of this nearby cluster have 
presented discordant results about the degree of dynamical mixing in 
M13's inner region. Using ground-based (\emph{UBVI}) photometry, we show 
that cumulative radial distributions of stars on the blue and 
red sides of the red giant branch are statistically identical.
Interestingly, these results are obtained using data from large-aperture,
ground-based telescopes as well as a more modestly-sized 
instrument, and both are in agreement with previous work done 
using $HST$ and Str\"omgren photometry. Results are derived using the 
\cubi\ index, shown to be sensitive to compositional differences.
We discuss our conclusions that the chemically distinct populations within 
M13 may be dynamically mixed in the context of published 
results from simulations.
\end{abstract}


\section{Introduction} \label{sec:intro}
Spectroscopic analysis of stars in Galactic globular clusters (GCs) has produced 
a paradigm for cluster evolution that includes the presence or formation of multiple 
stellar populations (MSPs). Such MSPs are chemically
different in light elements (e.g. He, C, N, O, Na) in ways that are typically explained in 
the context of advanced nucleosynthetic sequences occurring within a first generation of 
stars, where nuclear processed material is brought to the surface 
and dispersed into the intracluster medium via slow winds, rapid rotation, or some other
mechanism. According to this model, gas enriched in 
particular elements like N and Na and depleted in elements like C and O coalesces in 
the cluster core and forms a second, enriched generation of stars.
While this model does not neatly explain every GC's observed abundance pattern 
\citep[e.g.][]{bas15} or ratio of enriched to unenriched stars \citep{car10f}, 
it generally does a sufficiently adequate job of qualitatively explaining the vast majority 
of Milky Way GCs studied to-date that it has become the current leading model describing 
GC chemical evolution.

While the study of GC MSPs has largely been done using spectroscopy, given its 
advantage in identifying true chemical differences, photometric studies have lent themselves to expanding the study of GCs in this area as well. Using photometry to distinguish stellar subpopulations requires high resolution and S/N, and this is most easily accomplished using space-based or large ground-based observatories. 
Indeed, a series of work 
done by \citet{pio15} handily demonstrated the utility of the Hubble Space Telescope (hereafter \emph{HST}) in identifying 
these subpopulations from high-precision photometry. \citet{yon08}
demonstrated that ground-based photometry in intermediate-width Str\"omgren filters could be successfully used in distinguishing populations of stars within GCs by identifying subtle color 
differences among red giant branch (RGB) stars 
resulting from variations in atomic and/or molecular absorption within these passbands. 

Broadband Johnson \emph{U} and \emph{B} filters were used 
in conjunction with spectroscopy by \citet{mar08e} to show that differences in ($U-B$) color corresponded to differences in O, Na, and N abundances, where relatively Na-poor/CN-weak stars
lie on the blue side of the RGB and relatively Na-rich/CN-strong (``enriched'') stars lie on the red side, a result of 
the presence of several CN and NH absorption features in the $U$ passband. \citet{kra10a,kra10b} and \citet{kra11} further illustrated the distinction between these subpopulations on the RGB 
by revealing radial distribution differences between red RGB and blue RGB stars,
where the red RGB stars in their studies were reported to be more centrally concentrated. This 
finding was further emphasized by \citet{lar11} (hereafter referred to as \citetalias{lar11}), who looked at a collection of Sloan Digital Sky Survey \citep[SDSS; in particular][]{aba09} GCs 
and identified radial segregation in most of them using SDSS photometry alone. It remains a contested point whether segregated radial distributions are perhaps fairly common or relatively uncommon \citep[e.g.][]{lar15, lim16}. \citet{van15}, for example, reported that 80\% of their sample of 48 GCs appeared to be well-mixed.

Results for individual clusters are not always in agreement. The comprehensive study by \citetalias{lar11} of 
GCs in the SDSS database suggested that the Galactic GC M13 has a centralized concentration of enriched stars when examined over a radial range of 0.7 -- 6.7 half-light radii (\rh). This is seemingly at odds with the study by \citet{sav18} (hereafter referred to as \citetalias{sav18}), which used \emph{HST} data and ground-based Str\"omgren data from the Isaac Newton Telescope (hereafter INT) out to a radial distance of 6.5 \rh\ and reported M13 to be well-mixed over the entire radial range. Given the angular size and visual brightness of M13, this cluster is of fundamental value to the study of dynamical mixing among chemically distinct subpopulations because it should be among the best opportunities to obtain meaningful and reliable results using a large sample of stars that are well-resolved inside 2 \rh. In short, getting this cluster right is important for the validation of dynamical simulations. Furthermore, if the limiting factors in such work are 
resolution and S/N, combining the crowded-field analysis power of the DAOPHOT photometry suite \citep{ste87,1988AJ.....96..909S} with a significant amount of exposure time could potentially open up this space to sub-1-meter telescopes as well.

In this study we describe a procedure for attempting to distinguish stellar subpopulations in 
Galactic GCs using a modestly-sized telescope, and the application of that procedure to the Galactic GC 
M13. In $\S$ \ref{sec:data} we detail our data acquisition and reduction steps, and the membership selection procedure is described in $\S$ \ref{sec:proc}. The application of our procedure to M13 is detailed in $\S$ \ref{sec:valid}, where we also describe the rigorous testing we performed to give us confidence in our data and our procedure. In $\S$ \ref{sec:conflictres} we tackle the question of why previous results from \citetalias{lar11} and \citetalias{sav18} appear to be contradictory, and then we compare the result of this study with dynamical predictions in $\S$ \ref{sec:expectations}. Finally, we discuss our conclusions in $\S$ \ref{sec:concl}.

\section{Observations and Data Reduction}\label{sec:data}
Our data set for M13 was obtained over multiple nights between June 2016 and June 2019 using the Calvin University 0.4-m robotic observatory, located in Rehoboth, New Mexico. The data were collected using an SBIG ST-10XE thermoelectrically-cooled
CCD with a plate scale of 1.31'' per 2$\times$2-binned pixel, producing a field of view of roughly 24' $\times$ 16'. Images were taken in Johnson-Cousins UBVI. The total effective exposure times in the (\emph{U, B, V, I}) filters were (63 600, 23 400, 8 910, 8 820) seconds, respectively. All images were 
reduced via standard procedures using the MaxIm DL image acquisition, reduction, and analysis software package. All images passing an adopted seeing quality threshold of $\leq 3.9$ arcsec (3.0 pixels) FWHM in \emph{VI} and $\leq 4.7$ arcsec (3.6 pixels) FWHM in \emph{UB} were weight-averaged in IRAF\footnote{IRAF is distributed by the
National Optical Astronomy Observatory, which is operated by the Association of Universities for
Research in Astronomy (AURA) under a cooperative agreement with the National Science Foundation.}, where the
weighting factors were a function of each image's background sky noise. This resulted in a single master image in each filter with S/N $\geq 100$ along the portion of the RGB we targeted. The seeing threshold values were selected based on what was determined to be ``optimal'' seeing for our instrument and site.


Because of the crowded nature of these fields, the DAOPHOT/ALLSTAR photometry suite \citep{ste87,1988AJ.....96..909S} was employed to obtain instrumental photometry for stars in each master image. Briefly, this procedure involved carefully
constructing a PSF from a group of $\sim 50$ selected stars within each master image and using this PSF to fit and subtract detected stars in an interactive fashion, where each iteration completely re-fitted the original image using a list of all stars detected up to that point. The process of searching for and fitting stars in each image was iterated 3--4 times to obtain photometry that had acceptably low uncertainties. DAOMASTER produced our final cross-matched multiband catalog. DAOPHOT provides a \emph{sharp} statistic that quantifies the detection as being particularly point-like or extended, where the parameter is defined to be zero for point-like detections. Following the lead of \citetalias{sav18} and others, we adopted a cutoff value of $|sharp| \leq 0.3$ to exclude extended sources and artifacts. 

Our results rely on having photometry with very low photometric uncertainties, and the following procedure does not require that
the data be transformed to the standard Johnson-Cousins system. Therefore, to avoid introducing added error the instrumental magnitudes alone were considered sufficient.

Finally, differential reddening is an important consideration when closely examining cluster photometry. In their analysis of differential reddening for 66 Galactic GCs, \citet{bon13} reported that M13 exhibited minimal differential reddening ($\langle  \delta E(B-V)\rangle = 0.026 \pm 0.009 $). Given this, we opted to disregard differential reddening as a significant source
of dispersion among our M13 sample. 

\section{Cluster Membership Selection Procedure}\label{sec:proc} 
To clean the data sets of potential contamination and isolate our final sample of RGB stars, several cuts were instituted. These cuts involved 
using photometric uncertainty, radial distance from the cluster core, and position on a color-color diagram. Wishing not to bias our results via the quantitative choices we made in these cuts, we also sought to determine the degree to which the overall result was sensitive to some of the adopted cut levels. This validation is described in $\S$ \ref{subsec:paramtest}.

\begin{figure}
\plotone{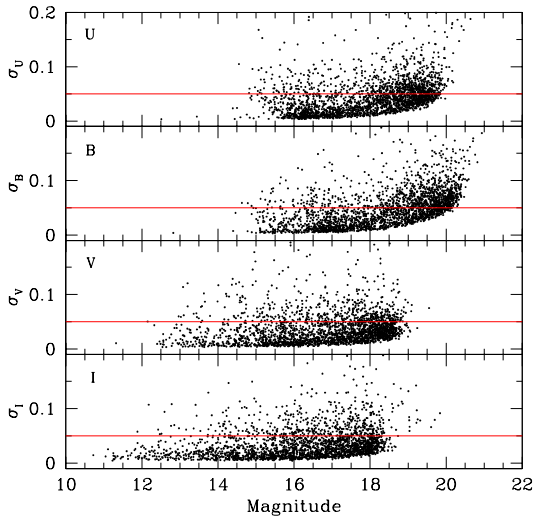}
\caption{The photometric uncertainty as a function of \emph{instrumental} magnitude for all stars in our sample. The horizontal red line indicates the cut at 0.05 mag we adopted as our standard cut.
\label{fig:errcut}}
\end{figure}

The first cut omitted stars with unacceptably large photometric uncertainty. The literature contains different philosophies on whether such photometric error cuts should be done using a fit to the uncertainty-magnitude plot \citep[e.g.][]{2013AJ....146...57C} or using a fixed value
\citep[e.g.][]{cle11}, and one can envision scenarios where one or the other approach could be preferred. 
The sensitivity of our analysis relies upon our using only stars with the best photometry, rather 
than photometry that is as good as could be expected for a given brightness, so by this rationale we implemented a
fixed value cut of 0.05 mag, illustrated for our four filters in Figure \ref{fig:errcut}. 

\begin{figure}
\plotone{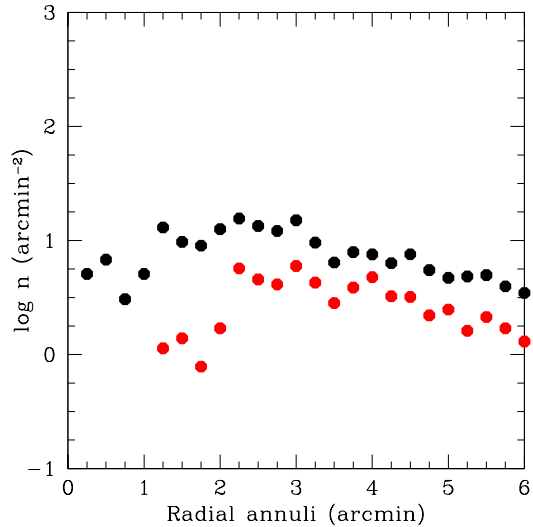}
\caption{The number density of stars (stars per $\rm{arcmin}^{2}$) in our M13 data set as a function of radial distance from the cluster center. Black data points represent all stars that passed only the uncertainty cut, while red data points represent only the stars that were subsequently used in our multiple population study described in Section \ref{sec:valid}. Completeness in our final sample begins to drop at a radial distance of approximately 2 arcmin.
\label{fig:numdens}}
\end{figure}

The second cut that was implemented involved radial distance from the cluster center, drawn from \citet{har96}. To this end, two cuts were implemented for M13: an inner radial cut to avoid stars deemed questionable due to blending, and an outer radial cut to minimize foreground/background contamination. Crowding in the cluster core has a significant impact on ground-based data, leading to a decrease in sample completeness. Figure \ref{fig:numdens} illustrates the effect of this crowding, where inside 2 arcmin from the cluster center the number density of stars begins to drop off instead of continuing to rise, indicating a decrease in the completeness of our sample. Given this observation, we adopted an inner radial cut of 2 arcmin, which corresponds to 1.2 \rh. An outer radius of 4 \rh\ was adopted due to the physical limitations of our CCD.

\begin{figure*}
\plotone{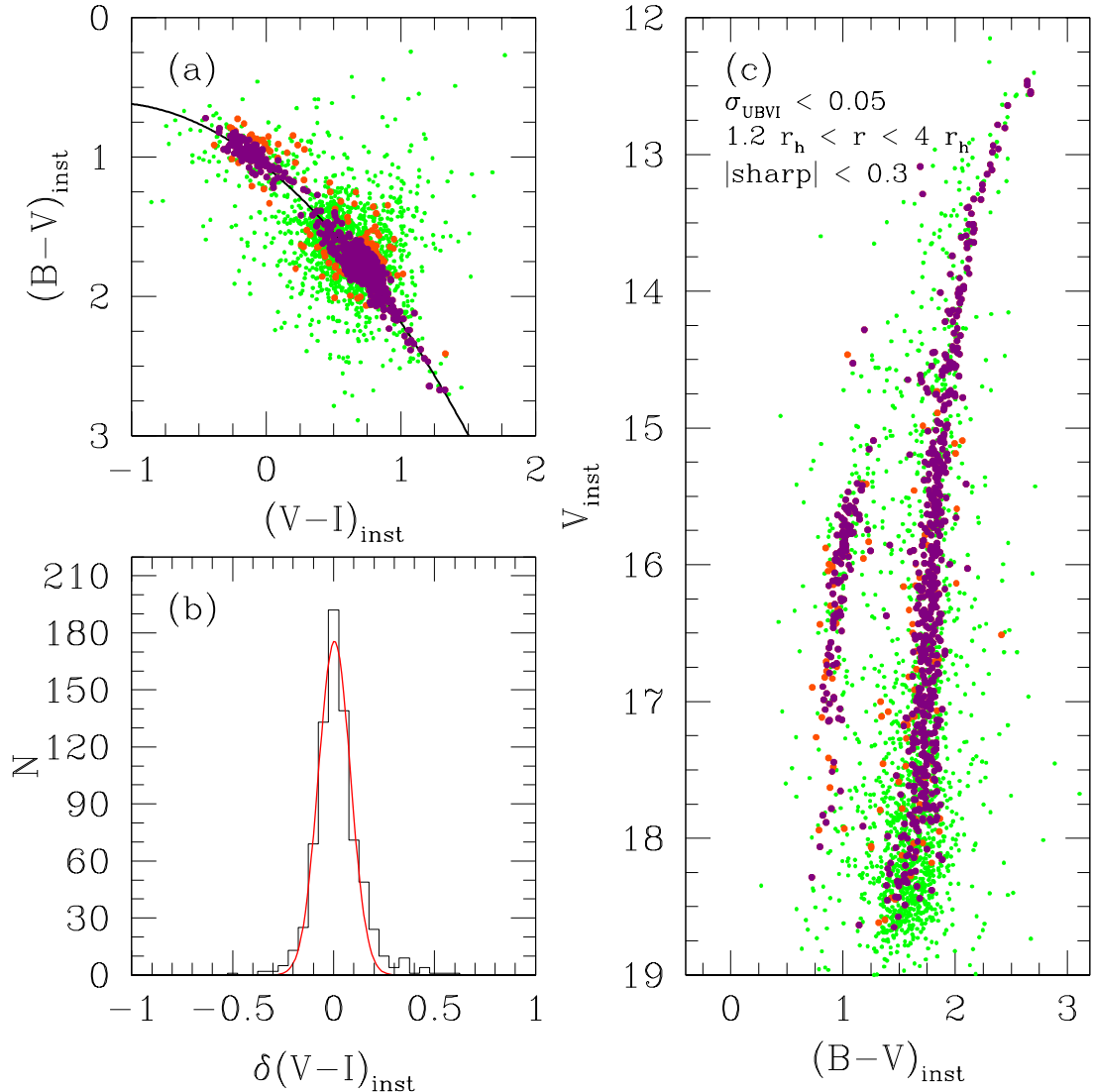}
\caption{$(a)$: The ($B-V$)-vs-($V-I$) color-color diagram for M13. The green points correspond to all stars detected in the field by DAOPHOT, while the orange points correspond to stars remaining after the uncertainty, $sharp$, and radial distance cuts described in the text. A polynomial was fit to these data and subtracted off, resulting in the quantity $\delta$($V-I$). $(b)$: The histogram of $\delta$($V-I)$. Stars within $2\sigma$ of the mean based on the Gaussian fit shown (red line) were retained and are plotted as purple points throughout this figure. $(c)$ The resulting color-magnitude diagram after this cut, where the purple points were the stars that passed the described cuts and defined our sample of adopted cluster members.
\label{fig:cmdrgb}}
\end{figure*}

The third cut that was implemented involved removing interloper stars along the RGB. In considering attempts to decontaminate the color-magnitude diagram (CMD) of interloper stars, we utilized the Besan\c con Galactic model of \citet{rob03} to statistically determine the expected number of stars in the direction of M13 that fell along the region of the RGB we ultimately analyzed. We found that the number of contaminating stars on this region of the CMD was minimal ($< 5$), so attempts to statistically decontaminate were omitted. However, omitting foreground/background contamination was necessary to more clearly resolve the RGB for later steps in our procedure. To do this we utilized a quadratic fit to the ($B-V$)-vs-($V-I$) color-color diagram, shown in Figure \ref{fig:cmdrgb}$a$. This color-color combination was chosen because it had both the smallest overall photometric uncertainty and the smallest amount of expected spread resulting from the presence of chemical composition differences among its RGB stars \citep{sbo11}. This fit was subtracted off from the data points and stars within 2 standard deviations from the mean $\delta$($V-I$) were retained. Results from this were subsequently compared to results derived from a cutoff at 1 standard deviation as well.

Stars that passed these cuts were adopted as our candidate cluster members.

\section{Distinguishing Multiple Populations}\label{sec:valid} 
\subsection{Review of Prior Methods}\label{subsec:priorproc}
The body of literature on identifying MSPs in GCs using ground-based photometry describes approaches that differ to varying degrees. In general, the process involves adopting a fiducial line to represent the shape of the RGB, and then assessing the position of RGB stars in color space with respect to this fiducial line. Some approaches fit such a line to the RGB \citep[e.g.][]{kra10b}, others drew such a ``ridge line'' in by hand through the RGB \citep[e.g.][hereafter referred to as \citetalias{mon13}]{mon13}, while still others defined a fiducial line along the edge of the RGB \citepalias{lar11}. In any case, the purpose of such a line was to remove the temperature effect from each star's color and defined a ``$\delta$(color)'' residual as the distance of each star from the adopted fiducial line due to intrinsic chemical differences. Stars at too great a distance from this line, such as horizontal branch (HB) or asymptotic giant branch (AGB) stars or non-cluster stars, were removed via some form of $\delta$(color) cut. This cut was identified by eye \citep[e.g.][]{kra10b} or statistically \citepalias[e.g.][]{mon13}.


Distinguishing stellar subgroups visually on a typical CMD is not easy. If spectroscopic data is available, then one can tag stars on a CMD based on their degree of enrichment in key light elements \citepalias[e.g. Figure 6 in][]{sav18}. In such instances, individual RGB loci appear largely separated by their degree of enrichment. However, this is not always convenient or possible. To address this, \citetalias{mon13} defined a new pseudo-color index \cubi\ = ($U-B$)$-$($B-I$), in the spirit of \citet{mil13}, to enhance the impacts that the ($U-B$) color has in revealing differences in light-element abundance \citep[e.g.][]{mar08e} and that ($B-I$) has in revealing He abundances \citep[e.g.][]{pio07}. On a CMD, this quantity typically reveals multiple loci in the RGB region. \citetalias{ste19} used the same \cubi\ definition and visually identified instances where a CMD using this pseudo-color index revealed substructure on the RGB. A similar use of a pseudo-color index was performed by \citetalias{sav18} with Str\"omgren photometry and the index $c_y = (u-v)-(v-b)-(b-y)$ \citep[introduced by][]{yon08}. This index also reveals chemical differences among stars on a CMD, in this case differences in N abundance, and results in multiple loci along the RGB that are nearly vertical in $y$-magnitude along much of the RGB, particularly the lower half. Dividing these multiple RGB loci was then done by drawing a line by hand or fitting a function.

RGB stars within some brightness range were then split into subgroups that were redder or bluer than some criterion. Typically, but not always, that dividing line was the fiducial/ridge line. Some groups \citep[e.g.][]{kra11} identified what we will call a ``\emph{fixed zone of avoidance}'' (ZOA) along the ridge line to omit stars whose uncertainties introduce ambiguity as to which subgroup they belong. Other studies disregarded this step. Cumulative radial distributions (CRDs) of the blue and red subgroups were then examined and quantified using a Kolmogorov-Smirnov (KS) test probability of the two CRDs being drawn from the same parent distribution.

\subsection{Our Approach}\label{subsec:ourproc}
\begin{figure}
\plotone{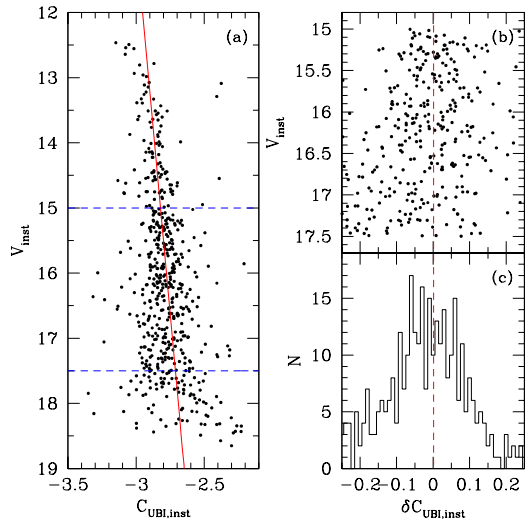}
\caption{$(a)$: $V_{inst}$ versus \cubi\ for our M13 candidate cluster members. The horizontal blue dashed lines represent the magnitude range used for subsequent steps, chosen to resemble the magnitude ranges used by other studies. The solid red line indicates a linear fit to all of the stars shown in this panel and serves to divide the RGB into two subgroups. $(b)$: The fit-subtracted CMD of all stars shown in (a), but restricted to highlight only the adopted magnitude range. Our data set does not reveal significant visual indication of distinct subgroup loci due to the relatively small sample size and the impact of scatter that is larger than other studies. $(c)$: The data from (b) in histogram form. There is some suggestion of a meaningful gap in the center that divides two subgroups. This feature was not significantly dependent on bin size and appeared even with bin sizes five times larger or half the size depicted here.
\label{fig:rgbdiv}}
\end{figure}

Figure \ref{fig:rgbdiv} illustrates our approach in a nutshell. We utilized the \cubi\ definition and examined a CMD showing $V$-magnitude as a function of \cubi. Similar to \citetalias{lar11} and \citetalias{sav18}, we isolated a 2.5-magnitude range along the middle region of the RGB and, noting from \citetalias{mon13} that this region was not exactly vertical, performed a linear fit to all candidate cluster members along the RGB. This fit served to act as our division between prospective subgoups. While \citetalias{mon13} claimed to have identified three subgroups in M13, we followed the simpler assumption of two subgroups: one primordial and one enriched in some way. To ascertain the impact of fitting this linear division along the full RGB, we performed a subsequent test fitting only within the 2.5-magnitude window as well. It is worth noting that this process assumes the existence of blue and red subgroups on the RGB, which for M13 has previously been demonstrated \citepalias[e.g.][]{mon13, sav18}.

Subtracting off this fit, removing HB stars using a simple color cut, and creating a histogram indicates that while there is no substructure visually apparent on the RGB (see Figure \ref{fig:rgbdiv}$b$), the histogram in Figure \ref{fig:rgbdiv}$c$ does provide hints of such overdensities, separated in $\delta$\cubi\ by an amount comparable to that seen by \citetalias{mon13}. Using the identification of \citetalias{mon13}, we adopted the left side of this histogram as the ``primordial'' group and the right side of this histogram as the ``enriched'' group. To avoid the impact of photometric uncertainty on whether a star gets placed in either group, we introduced what we refer to as a \emph{``dynamic zone of avoidance''}. In this technique, rather than adopting a fixed-width ZOA as \citet{kra11} did we allow each star's uncertainty in \cubi\ to dictate whether it be included or not. If a star's value of \cubi\ $\pm\ \sigma$(\cubi) placed it in a position where it could possibly lie on the other side of the midline, that star was omitted. 

\begin{figure*}
\plotone{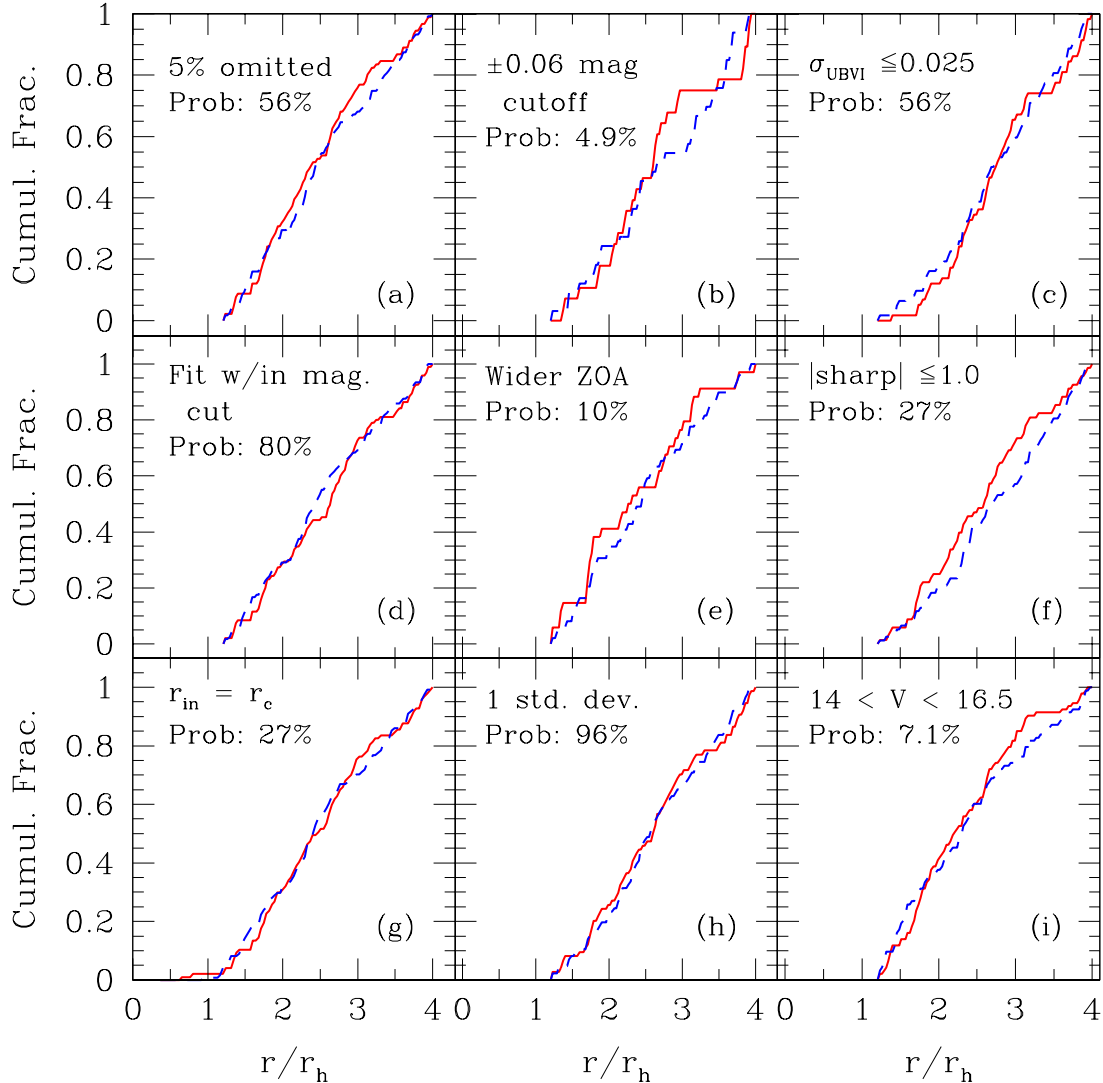}
\caption{$(a)$: The cumulative radial distribution for the two subgroups of stars selected in Figure \ref{fig:rgbdiv}. The blue (dashed) and red (solid) lines correspond to the subgroups associated with having primordial and enriched chemical abundances, respectively, as identified by \citetalias{mon13}. This represents our ``standard'' process. Panels (b) -- (i) are the same as (a) but with the following changes: $(b)$: Omitting RGB stars at a narrower cutoff range in $\delta$\cubi\ to reflect the RGB width reported by \citetalias{mon13}. 
$(c)$: Tighter photometric uncertainty cut.
$(d)$: Fitting the red line in Figure \ref{fig:rgbdiv}$a$ only within the adopted 15-17.5 magnitude range.
$(e)$: Wider zone of avoidance (ZOA).
$(f)$: More generous \emph{sharp} cutoff value.
$(g)$: Inner radial cut of 1 $r_c$.
$(h)$: Cut of 1 standard deviation from the mean fit to the ($B-V$)-vs-($V-I$) diagram.
$(i)$: Magnitude range shifted higher up along the RGB.
No meaningful difference between the two distributions is identified in any instance.
\label{fig:crds}}
\end{figure*}

\citetalias{mon13} used a limit on \cubi\ in order to further eliminate contamination or possible AGB stars by instituting a 5\% cut, where the 5\% reddest and bluest stars in $\delta$\cubi\ were omitted from consideration at this point. This produced a range in $\delta$\cubi\ of about 0.12 mag in their work. Our scatter is larger than this, where omitting the 5\% bluest and reddest stars introduces a cutoff at approximately $\delta$\cubi\ = $\pm0.2$ mag. Panel (a) of Figure \ref{fig:crds} illustrates the cumulative radial distributions (CRDs) of the blue (primordial) and red (enriched) subgroups from our sample using the selection process described above, omitting the reddest and bluest 5\% of stars. As can be seen, the two distributions do not appear to differ from one another, with the KS probability of being drawn from the same parent distribution at 56\%.  While we cannot make any claims about the presence or absence of radial concentration among the enriched population outside of 4 \rh\ using our data set alone, we do conclude that inside 4 \rh\ the two subgroups appear to be dynamically mixed.

\subsection{Testing the Impact of Parameter Selection}\label{subsec:paramtest}
The process of both producing a sample of candidate cluster members and isolating them into two distinct subgroups involves making certain quantitative choices. In this section we explore the effect of the particular choices we made by adjusting the values used and examining the impact of those adjustments on the CRDs.

We start by recalling the choice described at the end of $\S$ \ref{subsec:ourproc} where we adopted a limit on the width of the RGB in \cubi\ space by omitting the reddest and bluest 5\% of stars in our sample. This choice could arguably be supported by panel (c) of Figure \ref{fig:rgbdiv}, where the $\pm 0.2$ mag range this introduces does appear to reflect the width of the RGB as seen in our data. Admittedly, this width is much broader than the width observed by \citetalias{mon13}, so one question is whether or not our results depend on using the RGB width drawn from our own data set or that of \citetalias{mon13}. To test this, we adopted the width they reported, drawing a $\pm 0.06$ mag cutoff along the fit line. The CRDs associated with this test are shown in panel (b) of Figure \ref{fig:crds}, along with the corresponding KS probability. The result does not appear to depend on the choice of RGB width. 

One might question whether an uncertainty cut of 0.05 mag allows into the sample stars with photometric errors that are too large to be reliable. To this end, we explored the effect of decreasing the limit to 0.025 mag, shrinking the number of stars but retaining only those with the best measurements. Panel (c) in Figure \ref{fig:crds} illustrates the new CRDs, which are again statistically indistinguishable from one another. 

In Figure \ref{fig:rgbdiv} we showed the linear fit to the \cubi\ RGB using all stars in that sample. It is reasonable to check whether this is the best approach, given that the bright tip of the RGB exhibits a curve in the results reported by \citetalias{mon13}. For this reason, we tested a linear fit using only RGB stars in the magnitude range we adopted. The CRDs produced by the subgroups defined by this fit are shown in Figure \ref{fig:crds}$d$.

We also consider the impact of the possibility that our photometric uncertainties are underestimated. In such a case, the dynamic ZOA we allowed in our standard procedure may not be large enough to assure that stars on the red side of the midline are indeed redder than average. To accommodate this possibility, we explored widening the ZOA to $\pm0.1$ mag, $\sim10$ times our mean uncertainty. Figure \ref{fig:crds}$e$ illustrates the CRDs from such an adjustment. ZOA widths of $\pm0.05$ and $\pm0.075$ yielded identical results.

Panel (f) in Figure \ref{fig:crds} illustrates the impact of increasing our \emph{sharp} limit, allowing more stars into the sample which have slightly less ``star-like'' light profiles. The result is unchanged. Not shown here is an additional test we performed by shrinking the limit on \emph{sharp} to 0.2, restricting the sample to stars with even more star-like light profiles. Again, the result was not affected.

It is typical to consider making an inner radial cut with ground-based photometry, as seeing will introduce a limit to how well the innermost regions of GCs can be resolved. Given its relatively large angular size and brightness, resolving stars inside 1 \rh\ may be conceivable if the observing site and conditions are optimal along with a high S/N ratio. Our standard procedure described above adopted an inner radius cut of 1 \rh\ to align it with the studies we use for comparison, namely \citetalias{lar11} and \citetalias{sav18} who also adopted inner radial limits of $\approx 1$ \rh\ for their ground-based M13 data. However, \citetalias{lar11} actually uses a somewhat smaller inner radial cut of 0.7 \rh, which while not terribly different does probe deeper into the cluster center. Given this fact, and the significance of their conclusion, we explored what our own CRDs would look like if we too probed deeper toward the cluster center. Panel (g) in Figure \ref{fig:crds} indicates the resulting population distribution for an adopted inner radius cut of 1 $r_c$, which corresponds to approximately 0.4 \rh\ \citep{har96}. As can be seen, the other cuts we utilized remove most of the stars inside 1 \rh\ within our sample already so decreasing the inner radial limit does not impact our result. 

When performing the fit to the color-color diagram shown in Figure \ref{fig:cmdrgb}$a$ we accepted stars that fell within $2\sigma$ of the mean fit. To explore the impact of this choice, we ran a test restricting the window of acceptance to just $1\sigma$. The result is shown in Panel (h) of Figure \ref{fig:crds}, and it can be seen that the result is the same.

Finally, panel (i) in Figure \ref{fig:crds} addresses the possibility that we are avoiding a significant red population in the cluster that has higher luminosity than our bright limit. The bright limit in the \citetalias{lar11} sample from SDSS lies somewhat higher up on the RGB, so we tested our sample by shifting the 2.5-mag-wide brightness window up to span the range $14 \leq V_{inst} \leq 16.5$. Regardless of whether or not we opted to perform the linear fit to the RGB within this new magnitude range, the KS probability remained essentially unchanged.

The similarity of these diagrams and relatively high KS probabilities indicate that our result is independent of the parameter choices we made. We conclude that through rigorous testing of our cut parameter choices, we consistently obtain the result that inside 4 \rh\ M13 appears to be well-mixed.

\section{Resolving the Conflict}\label{sec:conflictres}
The radial distributions of subpopulations in M13 have been studied previously by \citetalias{lar11} and \citetalias{sav18}, with conflicting results. The procedures and data sets they used have both similarities and differences. The two ground-based data sets from both studies were obtained using comparable instruments, have comparable seeing limits and photometric errors, were both measured using the DAOPHOT suite, and span nearly identical radial distance ranges. It seems safe to conclude that the two data sets are of equal or comparable quality overall. 

Both procedures involved adopting a fiducial line that allowed the photometry to be separated into red and blue subgroups. Str\"omgren photometry using appropriate pseudo-color indices has been shown to serve as an excellent tool for separating out chemically different subpopulations, exceeding that of SDSS ($u-g$) alone as used by \citetalias{lar11}. On the other hand, broadband ($U-B$) has been shown to have some success and it is not obvious that ($u-g$) should suffer compared to this. It seems reasonable, then, to conclude that it ought to be \emph{possible} to separate chemically different subgroups using ($u-g$) if the procedure is applied carefully. 

The fundamental difference between the two procedures described by \citetalias{lar11} and \citetalias{sav18} is how photometric uncertainty was handled. \citetalias{sav18} proceeded with a cut on photometric uncertainty, rejecting stars outside three standard deviations from the mean uncertainty at a given magnitude in all filters. This has the effect of retaining stars that are measured as well as could be expected for their brightness. Although this could, in principle, lead to inclusion of stars with objectively large uncertainty, the magnitude range they ultimately used for the study omitted stars with that concern. 

On the other hand, \citetalias{lar11} did not cut by photometric uncertainty. Instead, once they had their color differentials $\Delta_{u-g}$ (the color difference between each star and the adopted fiducial line) they divided these values by the color uncertainty $\sigma_{u-g}$, with the assumption that this normalization accounts for the uncertainty contribution by ``dividing it out,'' producing a $\Delta'_{u-g}$ value that is presumably uncertainty-accommodated. 

\begin{figure}
\plotone{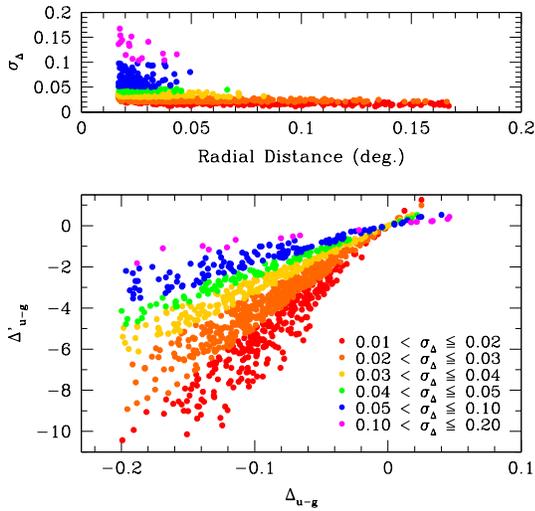}
\caption{\emph{Top panel}: Photometric uncertainty in the $\Delta_{u-g}$ color differential adopted by \citetalias{lar11} as a function of radial distance from the cluster center. \emph{Bottom panel}: The \emph{normalized} color differential $\Delta'_{u-g}$ derived by \citetalias{lar11} as a function of the original $\Delta_{u-g}$ value, colorized according to the photometric uncertainty in $\Delta_{u-g}$. While all values of $\sigma_{u-g}$ exist over the full range of $\Delta_{u-g}$, after the normalization the stars with larger $\sigma_{u-g}$ preferentially reside at redder values of $\Delta'_{u-g}$.
\label{fig:lardodeltaunc}}
\end{figure}

We reprocessed the \citetalias{lar11} data to investigate the impact of this approach using the procedural details the authors described as closely as possible. In the bottom panel of Figure \ref{fig:lardodeltaunc} we see that over the full range of the original $\Delta_{u-g}$ there are stars with small and large uncertainties, as expected. Once this normalization $\Delta'_{u-g} = \Delta_{u-g} / \sigma_{u-g}$ is performed, though, the stars with large photometric errors preferentially reside on the red end of the new $\Delta'_{u-g}$ distribution. As can be seen, stars with red $\Delta_{u-g}$ values remain on the red end of the $\Delta'_{u-g}$ distribution, regardless of their photometric uncertainty. ``Normalizing'' does nothing to allow these stars to shift to the blue side. Stars with blue $\Delta_{u-g}$ values can shift due to this normalization, and the effect is to preferentially move bluer stars with large uncertainties back toward the red end in $\Delta'_{u-g}$. The top panel of Figure \ref{fig:lardodeltaunc} shows that stars with the largest uncertainty in $\Delta_{u-g}$ reside closer to the cluster center. It seems possible to conclude, then, that the centrally concentrated red ($\Delta'_{u-g}$) subgroup identified by \citetalias{lar11} may have simply been the result of a bias in how the photometric uncertainty was addressed.

To verify that our suspicion is not unfounded, we applied our standard procedure described in $\S$ \ref{subsec:ourproc} to the \citetalias{lar11} M13 data drawn from SDSS, following the approach described by \citet{an08} with respect to determining an overall value of the DAOPHOT \emph{chi} and \emph{sharp} values for quality control purposes. Following our procedure required consideration of the \cubi\ index adopted earlier and how the SDSS filters compare. We are unaware of any rigorous attempts to test and define a similar index in the SDSS filter system. Given the spectral coverage of the SDSS filter set compared to the Johnson-Cousins system, it seems like a reasonable approximation to define an analogous pseudo-color $C_{u,g,i} = (u-g)-(g-i)$ and we will proceed using this index with the assumption that while it carries a degree of uncertainty about whether or not it is the \emph{most} appropriate pseudo-color in the SDSS filter set, it is unlikely to be the worst, nor likely to be entirely inappropriate for our purpose.

\begin{figure}
\plotone{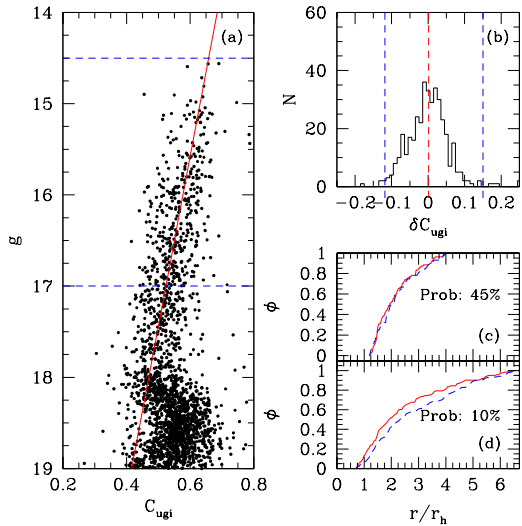}
\caption{Similar to Figure \ref{fig:rgbdiv}, where panels \emph{(a)} and \emph{(b)} correspond to panels \emph{(a)} and \emph{(c)} in Figure \ref{fig:rgbdiv}, respectively, but from the SDSS data set used by \citetalias{lar11} over the radial range of 1--4 \rh. Again, the horizontal blue dashed lines in $(a)$ indicate the magnitude range used, and the vertical blue dashed lines in $(b)$ indicate the cut made by omitting the extreme 5\% of the sample.
$(c)$ and $(d)$: CRDs for our process applied to the SDSS data set, with distance ranges of 1--4 \rh\ and 0.7--6.7 \rh, respectively. The central concentration vanishes in these CRDs, with KS tests being 35\% and 34\%, respectively. 
\label{fig:rgbdivlardo}}
\end{figure}

Figure \ref{fig:rgbdivlardo} illustrates the RGB of M13 in this SDSS pseudo-color. We again adopt a 2.5-magnitude-wide range, which is composed of the entire magnitude range \citetalias{lar11} adopts along with an additional half-magnitude on the bright end. We then examined CRDs over the radial range of 1.2--4 \rh\ (Figure \ref{fig:rgbdivlardo}$c$) for comparison with our results using other data sets, and over 0.7--6.7 \rh\ (Figure \ref{fig:rgbdivlardo}$d$), to compare with the results reported by \citetalias{lar11}. In both cases, the central concentration of red RGB stars appears to have vanished based on visual inspection as well as consideration of the KS probabilities, supporting the idea that how one chooses to address photometric uncertainty in this type of analysis is of paramount importance. 

\section{Dynamical Mixing Expectations}\label{sec:expectations}
Given the discrepancy of prior published results, examining the dynamical predictions should provide insightful expectation regarding the radial distribution of the two populations. Simulations of GC dynamics in the context of the formation of multiple populations are abundant, and generally these simulations predict that any subsequent generations of stars that form after the cluster initially forms should be centrally concentrated at first. The cluster then evolves through multiple dynamical mechanisms such that over time the new population(s) become increasingly mixed throughout the primordial population. 

\citet{dal19} reported on the results of N-body simulations that began with centrally concentrated second populations and thereafter evolved over a large number of dynamical timescales. During these simulations, they tracked the disparity between the radial distributions of the first and second populations, defining this difference using the parameter $A^+$, where

\begin{equation}
A^+ (R) = \int_{R_{min}}^{R} (\phi_{FP} (R') - \phi_{SP} (R'))dR'.
\end{equation}

This corresponds to the area enclosed between the CRDs of the two subpopulations, where $FP$ and $SP$ indicate the first (primordial) population and the second (enriched) population, respectively. A strongly centralized second population results in negative values of $A^+$, which they showed then evolve toward zero over time as the two populations mix. In their simulations, they only considered the innermost 2 \rh, thus dubbing their parameter $A_{2}^{+}$.

\begin{figure}
\plotone{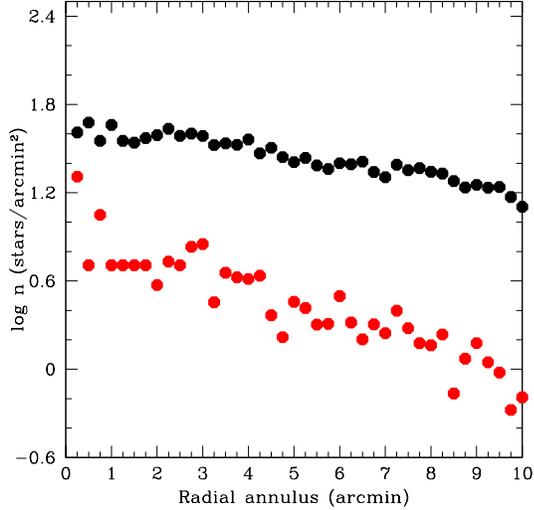}
\caption{Same as Figure \ref{fig:numdens}, but for the M13 photometry from \citetalias{ste19}.
\label{fig:numdensStetson}}
\end{figure}

\begin{figure}
\plotone{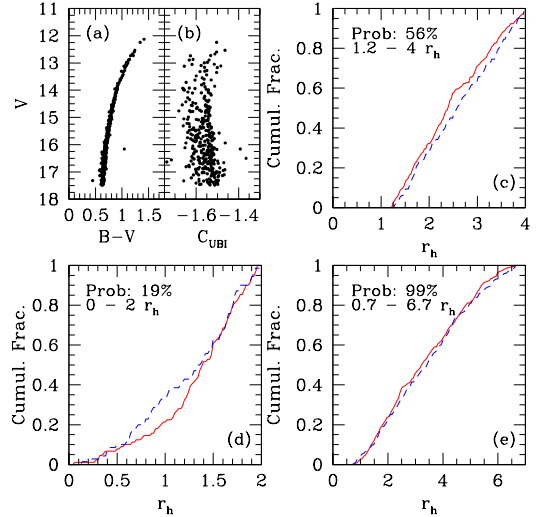}
\caption{$(a)$: CMD of M13 from \citetalias{ste19} illustrating the candidate cluster RGB stars only.
$(b)$: CMD of M13 shown in \cubi\ space.
$(c)$ -- $(e)$: CRDs of red (solid red line) and blue (dashed blue line) RGB stars over the radial distance ranges shown. KS probabilities for each CRD are shown, emphasizing the similarity between the two subgroups over each radial range.
\label{fig:cmdscrdsstetson}}
\end{figure}

While our own data set does not include the innermost 1 \rh, there is a data set that does. \citet{ste19}, hereafter referred to as \citetalias{ste19}, presented the public release of an extensive database of ground-based photometry for globular and open star clusters. We obtained the UBVI data set for M13 and
again examined the data set for completeness issues in the inner regions due to crowding. Figure \ref{fig:numdensStetson} indicates that beyond 2 arcmin the data appears to behave as expected, while inside 2 {arc-min} it is less obvious that the data set is complete. We applied the ``standard'' procedure described above to select the subset of candidate cluster members illustrated in panels (a) and (b) of Figure 
\ref{fig:cmdscrdsstetson}. We first examined the same radial distance range as we did for our own data set to determine whether or not our result depended on \emph{our data set}. Figure 
\ref{fig:cmdscrdsstetson}$c$ shows that this is not the case -- both our data set and the \citetalias{ste19} data set, undergoing the same procedure over the same radial range with all other cuts being the same, reveal what looks like a well-mixed distribution of stars.

Having established confidence in our procedure, we then turned our attention to the inner 2 \rh\ (3.4 arcmin) in the \citetalias{ste19} data set. We re-ran the selection and division process that extracted out the primordial and enriched subgroups within the 0 -- 2 \rh\ range this time (shown in Figure \ref{fig:cmdscrdsstetson}$d$). Again, we see (not surprisingly) that the two populations are generally well-mixed inside 2 \rh\ as well. Finally, we examined the radial range covered by \citetalias{mon13}, \citetalias{lar11}, and \citetalias{sav18} by using the \citetalias{ste19} archive photometry over the specific radial range used by \citetalias{lar11} and see, as shown in Figure \ref{fig:cmdscrdsstetson}$e$, that the cluster appears well-mixed throughout.

\begin{figure}
\plotone{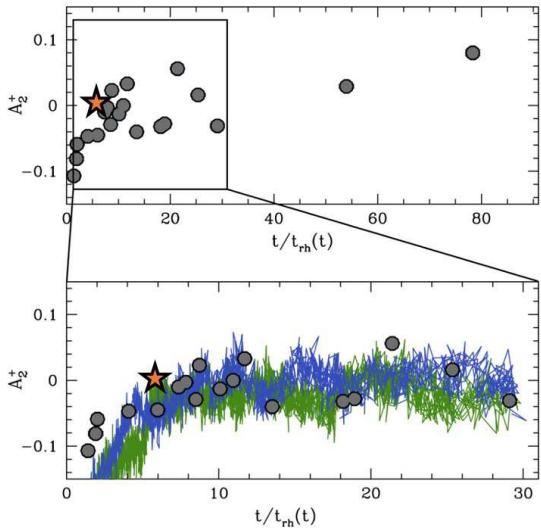}
\caption{A reproduction of Figure 3 from \citet{dal19}, illustrating the distribution of $A_{2}^{+}$ values drawn from their cluster sample (\emph{top panel}) and overlaid onto N-body simulations (\emph{bottom panel}). Our data point for M13, drawn from the analysis of the \citetalias{ste19} data, is overlaid onto both panels as an orange star. The placement of the orange star is consistent within the scatter illustrated by their simulations. (Reproduced by permission of the author and AAS.)
\label{fig:A+}}
\end{figure}

From the CRDs shown in panel (d) of Figure \ref{fig:cmdscrdsstetson}, we could potentially calculate a value of $A_{2}^{+}$ for comparison with the results from \citet{dal19}. Doing so depends on the data set being complete over this radial range or that the region where it is incomplete is sufficiently well-mixed that there is no difference between the radial distributions of the enriched and unenriched subgroups. While the number density of stars in Figure \ref{fig:numdensStetson} over this radial range does not smoothly follow the overall trend inside 2 arcmin, it also does not drop precipitously as our data did in Figure \ref{fig:numdens}. It is tantalizing enough that we pursued the option, arriving at a value of $A_{2}^{+} = 0.002$. Using M13's published age of 11.65 Gyr \citep[][and references therein]{for10} and the dynamical time of $t_{rh} = 1.995$ Gyr from \citet{har96}, we were able to place M13 on the simulation results of \citet{dal19}, shown in Figure \ref{fig:A+}. These results appear to illustrate that not only does M13 seem well-mixed inside 2 \rh, but that this observation appears consistent with dynamical predictions (notwithstanding our assumption that the data are relatively complete inside 2 arcmin). This supports the further observation from the \citetalias{ste19} data set, and our own, that M13 also appears to be well-mixed out to at least 6.7 \rh.

\section{Conclusion}\label{sec:concl}
In this work we addressed an apparent conflict in the literature between \citetalias{lar11} and \citetalias{sav18} regarding the nature of the radial distributions of chemically distinct subgroups within the Galactic GC M13. Visually distinguishing separate populations on the RGB is difficult for all but the best photometry, but it has been shown that the \cubi\ color index provides an optimal separation between the two populations. We proceeded to use this color index to cut our sample of stars down to stars very likely to be cluster members and divided that sample into known enriched and primordial subgroups guided by results of \citetalias{mon13} for M13. Cumulative radial distributions of the two subpopulations appear to be consistent with having been drawn from the same overall distribution. 

To demonstrate the veracity of our result we first rigorously tested our approach by selecting reasonable alternative values for our cut criteria. Notably, we considered the possibility that our data may be affected by underestimated photometric uncertainty by defining different ``zones of avoidance'' around the RGB midline, to assure ourselves that we were only considering stars that were truly redder or bluer than average. Second, we used archival data assembled by \citetalias{ste19} to assess whether our data set alone might be at all responsible for our result. Finally, we repeated our procedure on the SDSS data set used by \citetalias{lar11}. In all cases, cumulative radial distributions and KS probabilities supported the conclusion that M13 is well-mixed out to a radial distance of approximately 7 \rh\ from the cluster center. This observation using RGB stars supports the observation of \citet{van15}, in addition, who traced the populations using HB stars in M13 and also found them to be well-mixed.

We believe the source of the apparent conflict found in the literature regarding the dynamical state of M13 stems from the method \citetalias{lar11} adopts to account for the photometric uncertainty in the SDSS data set. The normalization process they describe appears to introduce a bias that preferentially moves stars with large uncertainties from the bluer side of the RGB to the redder side. Since stars with larger uncertainties are preferentially located near the center of the cluster, the result is an apparent central concentration of redder stars. In a future paper (in preparation), we will present the results of re-analysis of the other clusters in \citetalias{lar11}. Finally, the consistency among the results drawn from our own data, the \citetalias{ste19} archive, and the SDSS data also suggests that if there are no significant limitations to available exposure time, modestly-sized telescopes have the potential to make meaningful contributions to this type of research.

\acknowledgments
JPS, WBH, AJV, and ZJB acknowledge partial support from the Michigan Space Grant Consortium and the Calvin Research
Fellowship program. Students WBH, AJV, and ZJB are grateful for additional support from the Kanis, John Van Zytveld, and Hubert A. Vander Plas Memorial summer student research fellowships.
This research was also made possible by supporting funds from the Calvin University Science Division. Finally, we are indebted to Peter Stetson, Deokkeun An, Charles Kuehn, and Nathan De Lee for advice and guidance with DAOPHOT, and Charles Bonatto for providing us with a differential reddening map of M13 for reference. 

\software{DAOPHOT \citep{ste87}}
         

\bibliography{globs}{}

\begin{thebibliography}{}
\expandafter\ifx\csname natexlab\endcsname\relax\def\natexlab#1{#1}\fi
\providecommand{\url}[1]{\href{#1}{#1}}
\providecommand{\dodoi}[1]{doi:~\href{http://doi.org/#1}{\nolinkurl{#1}}}
\providecommand{\doeprint}[1]{\href{http://ascl.net/#1}{\nolinkurl{http://ascl.net/#1}}}
\providecommand{\doarXiv}[1]{\href{https://arxiv.org/abs/#1}{\nolinkurl{https://arxiv.org/abs/#1}}}

\bibitem[{{Abazajian} {et~al.}(2009){Abazajian}, {Adelman-McCarthy},
  {Ag{\"u}eros}, {Allam}, {Allende Prieto}, {An}, {Anderson}, {Anderson},
  {Annis}, {Bahcall}, {Bailer-Jones}, {Barentine}, {Bassett}, {Becker},
  {Beers}, {Bell}, {Belokurov}, {Berlind}, {Berman}, {Bernardi}, {Bickerton},
  {Bizyaev}, {Blakeslee}, {Blanton}, {Bochanski}, {Boroski}, {Brewington},
  {Brinchmann}, {Brinkmann}, {Brunner}, {Budav{\'a}ri}, {Carey}, {Carliles},
  {Carr}, {Castander}, {Cinabro}, {Connolly}, {Csabai}, {Cunha}, {Czarapata},
  {Davenport}, {de Haas}, {Dilday}, {Doi}, {Eisenstein}, {Evans}, {Evans},
  {Fan}, {Friedman}, {Frieman}, {Fukugita}, {G{\"a}nsicke}, {Gates},
  {Gillespie}, {Gilmore}, {Gonzalez}, {Gonzalez}, {Grebel}, {Gunn},
  {Gy{\"o}ry}, {Hall}, {Harding}, {Harris}, {Harvanek}, {Hawley}, {Hayes},
  {Heckman}, {Hendry}, {Hennessy}, {Hindsley}, {Hoblitt}, {Hogan}, {Hogg},
  {Holtzman}, {Hyde}, {Ichikawa}, {Ichikawa}, {Im}, {Ivezi{\'c}}, {Jester},
  {Jiang}, {Johnson}, {Jorgensen}, {Juri{\'c}}, {Kent}, {Kessler}, {Kleinman},
  {Knapp}, {Konishi}, {Kron}, {Krzesinski}, {Kuropatkin}, {Lampeitl},
  {Lebedeva}, {Lee}, {Lee}, {Leger}, {L{\'e}pine}, {Li}, {Lima}, {Lin}, {Long},
  {Loomis}, {Loveday}, {Lupton}, {Magnier}, {Malanushenko}, {Malanushenko},
  {Mandelbaum}, {Margon}, {Marriner}, {Mart{\'{\i}}nez-Delgado}, {Matsubara},
  {McGehee}, {McKay}, {Meiksin}, {Morrison}, {Mullally}, {Munn}, {Murphy},
  {Nash}, {Nebot}, {Neilsen}, {Newberg}, {Newman}, {Nichol}, {Nicinski},
  {Nieto-Santisteban}, {Nitta}, {Okamura}, {Oravetz}, {Ostriker}, {Owen},
  {Padmanabhan}, {Pan}, {Park}, {Pauls}, {Peoples}, {Percival}, {Pier}, {Pope},
  {Pourbaix}, {Price}, {Purger}, {Quinn}, {Raddick}, {Fiorentin}, {Richards},
  {Richmond}, {Riess}, {Rix}, {Rockosi}, {Sako}, {Schlegel}, {Schneider},
  {Scholz}, {Schreiber}, {Schwope}, {Seljak}, {Sesar}, {Sheldon}, {Shimasaku},
  {Sibley}, {Simmons}, {Sivarani}, {Smith}, {Smith}, {Smol{\v c}i{\'c}},
  {Snedden}, {Stebbins}, {Steinmetz}, {Stoughton}, {Strauss}, {Subba Rao},
  {Suto}, {Szalay}, {Szapudi}, {Szkody}, {Tanaka}, {Tegmark}, {Teodoro},
  {Thakar}, {Tremonti}, {Tucker}, {Uomoto}, {Vanden Berk}, {Vandenberg},
  {Vidrih}, {Vogeley}, {Voges}, {Vogt}, {Wadadekar}, {Watters}, {Weinberg},
  {West}, {White}, {Wilhite}, {Wonders}, {Yanny}, {Yocum}, {York}, {Zehavi},
  {Zibetti}, \& {Zucker}}]{aba09}
{Abazajian}, K.~N., {Adelman-McCarthy}, J.~K., {Ag{\"u}eros}, M.~A., {et~al.}
  2009, \apjs, 182, 543, \dodoi{10.1088/0067-0049/182/2/543}

\bibitem[{{An} {et~al.}(2008){An}, {Johnson}, {Clem}, {Yanny}, {Rockosi},
  {Morrison}, {Harding}, {Gunn}, {Allende Prieto}, {Beers}, {Cudworth},
  {Ivans}, {Ivezi{\'c}}, {Lee}, {Lupton}, {Bizyaev}, {Brewington},
  {Malanushenko}, {Malanushenko}, {Oravetz}, {Pan}, {Simmons}, {Snedden},
  {Watters}, \& {York}}]{an08}
{An}, D., {Johnson}, J.~A., {Clem}, J.~L., {et~al.} 2008, \apjs, 179, 326,
  \dodoi{10.1086/592090}

\bibitem[{{Bastian} {et~al.}(2015){Bastian}, {Cabrera-Ziri}, \&
  {Salaris}}]{bas15}
{Bastian}, N., {Cabrera-Ziri}, I., \& {Salaris}, M. 2015, \mnras, 449, 3333,
  \dodoi{10.1093/mnras/stv543}

\bibitem[{{Bonatto} {et~al.}(2013){Bonatto}, {Campos}, \& {Kepler}}]{bon13}
{Bonatto}, C., {Campos}, F., \& {Kepler}, S.~O. 2013, \mnras, 435, 263,
  \dodoi{10.1093/mnras/stt1304}

\bibitem[{{Carretta} {et~al.}(2010){Carretta}, {Bragaglia}, {Gratton},
  {Recio-Blanco}, {Lucatello}, {D'Orazi}, \& {Cassisi}}]{car10f}
{Carretta}, E., {Bragaglia}, A., {Gratton}, R.~G., {et~al.} 2010, \aap, 516,
  A55, \dodoi{10.1051/0004-6361/200913451}

\bibitem[{{Clem} {et~al.}(2011){Clem}, {Landolt}, {Hoard}, \&
  {Wachter}}]{cle11}
{Clem}, J.~L., {Landolt}, A.~U., {Hoard}, D.~W., \& {Wachter}, S. 2011, \aj,
  141, 115, \dodoi{10.1088/0004-6256/141/4/115}

\bibitem[{{Contreras Pe{\~n}a} {et~al.}(2013){Contreras Pe{\~n}a}, {Catelan},
  {Grundahl}, {Stephens}, \& {Smith}}]{2013AJ....146...57C}
{Contreras Pe{\~n}a}, C., {Catelan}, M., {Grundahl}, F., {Stephens}, A.~W., \&
  {Smith}, H.~A. 2013, \aj, 146, 57, \dodoi{10.1088/0004-6256/146/3/57}

\bibitem[{{Dalessandro} {et~al.}(2019){Dalessandro}, {Cadelano}, {Vesperini},
  {Martocchia}, {Ferraro}, {Lanzoni}, {Bastian}, {Hong}, \& {Sanna}}]{dal19}
{Dalessandro}, E., {Cadelano}, M., {Vesperini}, E., {et~al.} 2019, \apjl, 884,
  L24, \dodoi{10.3847/2041-8213/ab45f7}

\bibitem[{{Forbes} \& {Bridges}(2010)}]{for10}
{Forbes}, D.~A., \& {Bridges}, T. 2010, \mnras, 404, 1203,
  \dodoi{10.1111/j.1365-2966.2010.16373.x}

\bibitem[{{Harris}(1996)}]{har96}
{Harris}, W.~E. 1996, \aj, 112, 1487, \dodoi{10.1086/118116}

\bibitem[{{Kravtsov} {et~al.}(2010{\natexlab{a}}){Kravtsov}, {Alca{\'{\i}}no},
  {Marconi}, \& {Alvarado}}]{kra10a}
{Kravtsov}, V., {Alca{\'{\i}}no}, G., {Marconi}, G., \& {Alvarado}, F.
  2010{\natexlab{a}}, \aap, 516, A23+, \dodoi{10.1051/0004-6361/200913449}

\bibitem[{{Kravtsov} {et~al.}(2010{\natexlab{b}}){Kravtsov}, {Alca{\'{\i}}no},
  {Marconi}, \& {Alvarado}}]{kra10b}
---. 2010{\natexlab{b}}, \aap, 512, L6+, \dodoi{10.1051/0004-6361/200913749}

\bibitem[{{Kravtsov} {et~al.}(2011){Kravtsov}, {Alca{\'{\i}}no}, {Marconi}, \&
  {Alvarado}}]{kra11}
---. 2011, \aap, 527, L9+, \dodoi{10.1051/0004-6361/201015975}

\bibitem[{{Lardo} {et~al.}(2011){Lardo}, {Bellazzini}, {Pancino}, {Carretta},
  {Bragaglia}, \& {Dalessandro}}]{lar11}
{Lardo}, C., {Bellazzini}, M., {Pancino}, E., {et~al.} 2011, \aap, 525, A114+,
  \dodoi{10.1051/0004-6361/201015662}

\bibitem[{{Larsen} {et~al.}(2015){Larsen}, {Baumgardt}, {Bastian}, {Brodie},
  {Grundahl}, \& {Strader}}]{lar15}
{Larsen}, S.~S., {Baumgardt}, H., {Bastian}, N., {et~al.} 2015, \apj, 804, 71,
  \dodoi{10.1088/0004-637X/804/1/71}

\bibitem[{{Lim} {et~al.}(2016){Lim}, {Lee}, {Pasquato}, {Han}, \&
  {Roh}}]{lim16}
{Lim}, D., {Lee}, Y.-W., {Pasquato}, M., {Han}, S.-I., \& {Roh}, D.-G. 2016,
  \apj, 832, 99, \dodoi{10.3847/0004-637X/832/2/99}

\bibitem[{{Marino} {et~al.}(2008){Marino}, {Villanova}, {Piotto}, {Milone},
  {Momany}, {Bedin}, \& {Medling}}]{mar08e}
{Marino}, A.~F., {Villanova}, S., {Piotto}, G., {et~al.} 2008, \aap, 490, 625,
  \dodoi{10.1051/0004-6361:200810389}

\bibitem[{{Milone} {et~al.}(2013){Milone}, {Marino}, {Piotto}, {Bedin},
  {Anderson}, {Aparicio}, {Bellini}, {Cassisi}, {D'Antona}, {Grundahl},
  {Monelli}, \& {Yong}}]{mil13}
{Milone}, A.~P., {Marino}, A.~F., {Piotto}, G., {et~al.} 2013, \apj, 767, 120,
  \dodoi{10.1088/0004-637X/767/2/120}

\bibitem[{{Monelli} {et~al.}(2013){Monelli}, {Milone}, {Stetson}, {Marino},
  {Cassisi}, {del Pino Molina}, {Salaris}, {Aparicio}, {Asplund}, {Grundahl},
  {Piotto}, {Weiss}, {Carrera}, {Cebri{\'a}n}, {Murabito}, {Pietrinferni}, \&
  {Sbordone}}]{mon13}
{Monelli}, M., {Milone}, A.~P., {Stetson}, P.~B., {et~al.} 2013, \mnras, 431,
  2126, \dodoi{10.1093/mnras/stt273}

\bibitem[{{Piotto} {et~al.}(2007){Piotto}, {Bedin}, {Anderson}, {King},
  {Cassisi}, {Milone}, {Villanova}, {Pietrinferni}, \& {Renzini}}]{pio07}
{Piotto}, G., {Bedin}, L.~R., {Anderson}, J., {et~al.} 2007, \apjl, 661, L53,
  \dodoi{10.1086/518503}

\bibitem[{{Piotto} {et~al.}(2015){Piotto}, {Milone}, {Bedin}, {Anderson},
  {King}, {Marino}, {Nardiello}, {Aparicio}, {Barbuy}, {Bellini}, {Brown},
  {Cassisi}, {Cool}, {Cunial}, {Dalessandro}, {D'Antona}, {Ferraro}, {Hidalgo},
  {Lanzoni}, {Monelli}, {Ortolani}, {Renzini}, {Salaris}, {Sarajedini}, {van
  der Marel}, {Vesperini}, \& {Zoccali}}]{pio15}
{Piotto}, G., {Milone}, A.~P., {Bedin}, L.~R., {et~al.} 2015, \aj, 149, 91,
  \dodoi{10.1088/0004-6256/149/3/91}

\bibitem[{{Robin} {et~al.}(2003){Robin}, {Reyl{\'e}}, {Derri{\`e}re}, \&
  {Picaud}}]{rob03}
{Robin}, A.~C., {Reyl{\'e}}, C., {Derri{\`e}re}, S., \& {Picaud}, S. 2003,
  \aap, 409, 523, \dodoi{10.1051/0004-6361:20031117}

\bibitem[{{Savino} {et~al.}(2018){Savino}, {Massari}, {Bragaglia},
  {Dalessandro}, \& {Tolstoy}}]{sav18}
{Savino}, A., {Massari}, D., {Bragaglia}, A., {Dalessandro}, E., \& {Tolstoy},
  E. 2018, \mnras, 474, 4438

\bibitem[{{Sbordone} {et~al.}(2011){Sbordone}, {Salaris}, {Weiss}, \&
  {Cassisi}}]{sbo11}
{Sbordone}, L., {Salaris}, M., {Weiss}, A., \& {Cassisi}, S. 2011, ArXiv
  e-prints.
\newblock \doarXiv{1103.5863}

\bibitem[{{Stetson}(1987)}]{ste87}
{Stetson}, P.~B. 1987, \pasp, 99, 191, \dodoi{10.1086/131977}

\bibitem[{{Stetson} \& {Harris}(1988)}]{1988AJ.....96..909S}
{Stetson}, P.~B., \& {Harris}, W.~E. 1988, \aj, 96, 909, \dodoi{10.1086/114856}

\bibitem[{{Stetson} {et~al.}(2019){Stetson}, {Pancino}, {Zocchi}, {Sanna}, \&
  {Monelli}}]{ste19}
{Stetson}, P.~B., {Pancino}, E., {Zocchi}, A., {Sanna}, N., \& {Monelli}, M.
  2019, \mnras, 485, 3042, \dodoi{10.1093/mnras/stz585}

\bibitem[{{Vanderbeke} {et~al.}(2015){Vanderbeke}, {De Propris}, {De Rijcke},
  {Baes}, {West}, {Alonso-Garc{\'\i}a}, \& {Kunder}}]{van15}
{Vanderbeke}, J., {De Propris}, R., {De Rijcke}, S., {et~al.} 2015, \mnras,
  451, 275, \dodoi{10.1093/mnras/stv928}

\bibitem[{{Yong} {et~al.}(2008){Yong}, {Grundahl}, {Johnson}, \&
  {Asplund}}]{yon08}
{Yong}, D., {Grundahl}, F., {Johnson}, J.~A., \& {Asplund}, M. 2008, \apj, 684,
  1159, \dodoi{10.1086/590658}

\end{thebibliography}
\bibliographystyle{aasjournal}


\end{document}